# Benefits of direct electron detection and PCA for EELS investigation of organic photovoltaics materials

*Georg Haberfehlner[a,*], Sebastian F. Hoefler[b], Thomas Rath[b], Gregor Trimmel[b], Gerald Kothleitner[a,c], Ferdinand Hofer[a,c]*

[a] Institute of Electron Microscopy and Nanoanalysis, NAWI Graz, Graz University of Technology, Steyrergasse 17, 8010 Graz, Austria

[b] Institute for Chemistry and Technology of Materials (ICTM), NAWI Graz, Graz University of Technology, Stremayrgasse 9, 8010 Graz, Austria

[c] Graz Centre for Electron Microscopy, Steyrergasse 17, Graz, 8010, Austria

* Corresponding author: georg.haberfehlner@felmi-zfe.at

**Highlights**

- The performance of direct electron detection and conventional CCD imaging for EELS is compared
- Direct electron detectors reveal spectral features clearly
- Principal components analysis is essential to reveal donor/acceptor separation in PTB7-Th/O-IDTBR blend

**Abstract**

Electron energy-loss spectroscopy (EELS) is a powerful tool for imaging chemical variations at the nanoscale. Here, we investigate a polymer/organic small molecule-blend used as absorber layer in an organic solar cell and employ EELS for distinguishing polymer donor and small molecule acceptor domains in the nanostructured blend based on elemental maps of light elements, such as nitrogen, sulfur or fluorine. Especially for beam sensitive samples, the electron dose needs to be limited, therefore optimized acquisition and data processing strategies are required. We compare data acquired on a post-column energy filter with a direct electron detection camera to data from a conventional CCD camera on the same filter and we investigate the impact of statistical data processing methods (principal components analysis, PCA) on acquired spectra and elemental maps extracted from spectrum images. Our work shows, that the quality of spectra on a direct electron detection camera is far superior to conventional CCD imaging, and thereby allows clear identification of ionization edges and the fine structure of these edges. For the quality of the elemental maps, the application of PCA is essential to allow a clear separation between the donor and acceptor phase in the bulk heterojunction absorber layer of a non-fullerene organic solar cell.

**Keywords:** electron energy-loss spectroscopy; direct electron detection; principal components analysis; organic solar cell

**Introduction**

Electron energy-loss spectroscopy (EELS) in a scanning transmission electron microscope (STEM) provides a large amount of information about a sample at the nano- and atomic scale, such as elemental composition and bonding, or electrical and optical materials properties (Egerton, 2011). For elemental mapping, and especially for light elements, the high collection efficiency is its main advantage compared to energy-dispersive X-ray spectroscopy (EDS), while data processing and reliable extraction of element-specific signals can be challenging due to the overlay with the background signal and plural scattering effects.

An important limit for TEM investigations is the electron dose, which can change the material during the investigation. For many materials, including polymers, working at cryogenic temperatures is a possibility to increase the stability of samples, thereby allowing higher electron doses (Hofer et al., 1986). Equally important is the efficient usage of the available dose, which is possible by using new types of electron detectors.

In EELS usually a two-dimensional camera is used for recording of individual spectra, where the energy-axis is spread along one dimension of the camera, while pixels along the second dimension are generally summed up. Most commonly used are CCD or CMOS cameras detecting photons, with preceding scintillator and fibre optics converting electrons to photons (Mooney et al., 1990). More recently direct electron detectors have been introduced in TEM, which have shown their benefits in particular for low-dose imaging in life science applications such as single particle analysis (Bammes et al., 2012; De Zorzi et al., 2016; McMullan et al., 2014; Ruskin et al., 2013; Wu et al., 2016). By eliminating the scintillator/fibre optic stack direct electron detectors significantly reduce noise. Using the so-called counting mode of a direct electron detection camera, single electrons can be detected at each pixel, if no more than a single electron arrives at each detector pixel at each frame. The recordable dose is therefore limited by the frame rate of the camera, e.g. 400 frames per second for the Gatan K2 camera (Li et al., 2013). Thereby, if the dose rate is low enough, noise is essentially limited to shot noise. Additionally, due to the low thickness of direct electron detectors, cross-talk between neighboring pixels is almost completely eliminated. Direct electron detectors can also be used as cameras for EELS, in order to limit the dose, and in addition the reduced cross-talk is also useful for optimal energy-resolution also for larger dispersions (Maigné and Wolf, 2018). The reduced noise clearly is an advantage in itself, but especially when using statistical data analysis methods, such as principal components analysis (PCA) (de la Peña et al., 2011; Jolliffe and Cadima, 2016; Spiegelberg and Rusz, 2017), the knowledge about the Poisson character of the noise allows efficient usage of such algorithms (Keenan and Kotula, 2004).

In this paper, we show a comparative study of EELS spectrum imaging using a conventional CCD (Gatan US1000) and direct electron detection camera (Gatan K2) on a polymer/small molecule blend film, analyzing the extraction of different ionization edges and elemental maps with and without the application of PCA. The investigated polymer/small molecule (PTB7-Th/O-IDTBR) blend is used as absorber layer of a non-fullerene organic solar cell. Photovoltaic properties of this material combination, phase separation in the polymer/small molecule absorber layer and its impact on device performance has been discussed in previous work (Hoefler et al., 2018, 2019a, 2019b), while in this paper the focus is on the EELS methodology and on the comparison of conventional CCD imaging and direct electron detection for the characterization of the phase separation in this blend. Generally, the electron microscopic investigation of the phase separation in polymer/small molecule absorber layer blends of non-fullerene organic solar cells is challenging due to the very similar chemical composition of both materials in the blend. For the investigated material combination, it was possible to distinguish both phases by spectral imaging based on the fact that only the conjugated polymer PTB7-Th contains fluorine and only the small molecule O-IDTBR contains nitrogen in the chemical structure (see Figure 1).

**Materials and Methods**

The investigated polymer/small molecule blend consists of a combination of the low bandgap conjugated polymer PTB7-Th (poly[4,8-bis(5-(2-ethylhexyl)thiophen-2-yl)benzo[1,2-*b*;4,5-*b'*]dithiophene-2,6-diyl-*alt*-(4-(2-ethylhexyl)-3-fluorothieno[3,4-*b*]thiophene-)-2-carboxylate-2-6-diyl]) as polymer donor blended together with the small molecule acceptor O-IDTBR ((5Z,5′Z)-5,5′-(((4,4,9,9-tetraoctyl-4,9-dihydro-*s*-indaceno[1,2-*b*:5,6-*b'*]dithiophene-2,7-diyl)bis(benzo[*c*][1,2,5]thiadiazole-7,4-diyl))bis(methanylylidene))bis(3-ethyl-2-thioxothiazolidin-4-one)), as shown in Fig. 1. Both materials were purchased from 1-Material. For the TEM investigations a blend with a donor:acceptor weight ratio of 1:1.5 was deposited by spin coating using a solution in *ortho*-dichlorobenzene onto a sodium chloride crystal, which was afterwards dissolved in water for the transfer of the polymer film to a TEM grid. Spin coating parameters were chosen to achieve a film thickness of ~80-90 nm (Hoefler et al., 2019b).

STEM investigations were performed in a FEI Titan³ G2 60-300, operated at 300 kV. The beam current was ~100 pA, this kept the dose rate below ~10 electrons/pixel/s within the energy range recorded on the K2 camera, so that mostly not more than one electron arrives at a single pixel within one frame. The beam semi-convergence angle was 19.6 mrad, the collection angle for EELS was 24.2 mrad using a 5 mm GIF entrance aperture. EELS spectrum images were acquired with a size of 182×171 pixels at a pixel size of 2 nm and with a dwell time of 10 ms per pixel. For acquisition with the K2 camera the used dispersion was 0.25 eV/ch, for acquisition of a spectral range from 120 eV to 1050 eV. For the CCD camera a dispersion of 0.5 eV/ch was used in order to span the same spectral range. For the CCD acquisition 130 times binning in the non-dispersive direction on the camera was used along with the high-quality readout of the CCD. For the K2 camera, binning is not applicable.

**Results and Discussion**

For a conventional CCD camera, the total noise is given by four components: shot noise, photon conversion noise, readout noise and gain noise (Egerton, 2011). For a direct electron detector in counting mode only shot noise remains, given by $N_S = \frac{\sqrt{N}}{s}$, where $N$ is the number of events in a single pixel and $s$ is a factor accounting for possible signal spread to neighboring detector pixels.

To illustrate the differences in noise and cross-talk between neighboring pixels, Fig. 2 shows a comparison of spectra acquired with a K2 camera compared to spectra acquired using a conventional Gatan US1000 CCD camera. Spectrum images on the CCD camera were acquired under the same conditions as on the K2 camera, but with a two times lower dispersion, to span the same energy range with the lower number of pixels of the CCD (2048 channels).

Fig. 2a shows a spectrum summed over a full spectrum image, where ionization edges for S, C, N and O can be clearly distinguished from the background. Only F cannot be clearly identified, due to its very low concentration. By contrast, in the signal acquired on a conventional CCD (Fig. 2b) separation of ionization edges from the background is much more problematic, especially due to systematic noise in the spectra, which leads to signal variations comparable to changes stemming from ionization edges of elements present in low concentration (e.g. N). Additionally, also the improved point spread function can be observed in the comparison, for example for the π*-peak in the C K-edge, which comes out clearly in the spectrum acquired on the K2 camera.

Spectra from single pixels within the spectrum image show significant noise, also for the K2 camera due to the presence of shot noise for low number of counts (Fig. 2c, green curve). The impact of Poisson noise is also visible for the CCD camera (Fig 2d, green curve), though it is smoothed out due to cross-talk between neighboring pixels and as the pixel size is reduced by a factor of 2 for the CCD camera, leading to a lower number of spectral channels.

To reduce the noise in single pixel spectra, principal components analysis (PCA) was performed on the spectrum images using the software Hyperspy (de la Peña et al., 2018) and taking into account the Poisson statistics of the noise. The scree plots for the spectrum images acquired with the K2 and the CCD camera are shown in Fig. 3, along with the loadings and factors of the first three principal components, which are selected, while all other components are discarded for denoising of the data.

The impact of PCA on spectra for both the K2 camera and the CCD is shown in Fig. 2, where red curves show PCA treated spectra. The removal of shot noise is apparent, which leads to very clear peaks in the single pixel spectrum on the K2 camera, while for the CCD camera systematic noise is still present after PCA.

Finally, elemental maps of the elements present in the sample are extracted from ionization edges (S-$L_{23}$; C-K, N-K, O-K, F-K) of both the K2 and CCD datasets with and without PCA treatment (see Fig. 4). The extraction of elemental maps is done using a fitting approach for each ionization edge, where a power-law background is fitted along with a Hartree-Slater model for the ionization edge, over the region around the onset of the ionization edge. Without PCA treatment, in both datasets features within the elemental maps are hardly visible. Only in the maps of sulfur and carbon, contrast changes can be observed, which can be attributed to sample features, but appear very noisy. While the contrast variations in the carbon maps correspond to changes in the HAADF signal, and can be thereby attributed to thickness variations, the sulfur maps show additional features with sizes in the range of few tens of nanometers, which can be attributed to the phase separation between PTB7-Th and O-IDTBR. In the PCA treated maps, noise is significantly reduced, and features are also visible in maps of other elements. Oxygen variations are correlated to the carbon map and thereby mostly represent thickness. The nitrogen maps mostly show an inverted contrast compared to the sulfur map. Nitrogen is only present in O-IDTBR and thereby the nitrogen-rich regions can be attributed to this material, while PTB7-Th shows a higher sulfur contrast. In the CCD data, the fluorine signal cannot be extracted, in the K2 map, the data is also very noisy, but contrast correlating with the sulfur map can be observed, which could be attributed to the single fluorine atom present in the PTB7-Th repeating unit. Mapping of fluorine can be improved by longer acquisition time, as shown previously in (Hoefler et al., 2019b). Fig. 5 shows maps of the relative composition of sulfur and nitrogen (calculated relative to all other elements present in the spectrum) (Hofer et al., 1997), where the two phases can be clearly separated.

Generally, a large improvement in the quality of elemental maps can be observed due to PCA for both datasets, while the differences between CCD and K2 data in the elemental maps are small. A likely reason for this is that direct electron detection mostly reduces systematic noise, which does not severely impact the extraction of elemental maps, as each spectrum in the spectrum image is affected in the same way. The fitting parameters of different spectra in a spectrum image, which determine the values in the elemental maps are all shifted by a similar value. Relative changes between the fitting parameters, which are visible in the elemental maps, are therefore not affected by systematic noise.

**Conclusion**

In this work, we investigated the impact of direct electron detection and of PCA on spectra and elemental maps from EELS spectrum image data on a polymer/small molecule blend sample. Direct electron detection provides major advantages in terms of quality of spectra, as systematic noise is suppressed, thereby allowing clear identification of ionization edges along with detailed fine structure information. PCA is essential for extracting elemental maps of sulfur and nitrogen in the dataset, which allows clear identification of the donor and acceptor regions in the polymer blend. This provides very good contrast between the two phases, despite very low differences in the chemical composition.

**Acknowledgments**

This work was carried out within the project "SolaBat – Solar cell meets battery – Realization of a hybrid energy system" funded by the Austrian "Climate and Energy Fund" within the program Energy Emission Austria (FFG No. 853 627). The support of Ilie Hanzu is gratefully acknowledged. The purchase of the K2-camera was supported by the "Zukunftsfonds Steiermark". Furthermore, this project has received funding from the European Union's Horizon 2020 research and innovation programme under grant agreement No. 823717 – ESTEEM3.

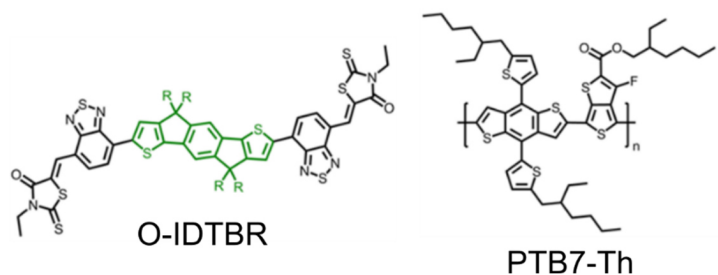

Fig. 1. Chemical structures of O-IDTBR (R = *n*-octyl) $C_{72}H_{88}N_6O_2S_8$ (wt.%: C: 65.1, H: 6.7, N: 6.3, O: 2.4, S: 19.3) and PTB7-Th $(C_{49}H_{57}FO_2S_6)_n$ (wt.%: C: 66.2, H: 6.5, F: 2.1, O: 3.6, S: 21.6)

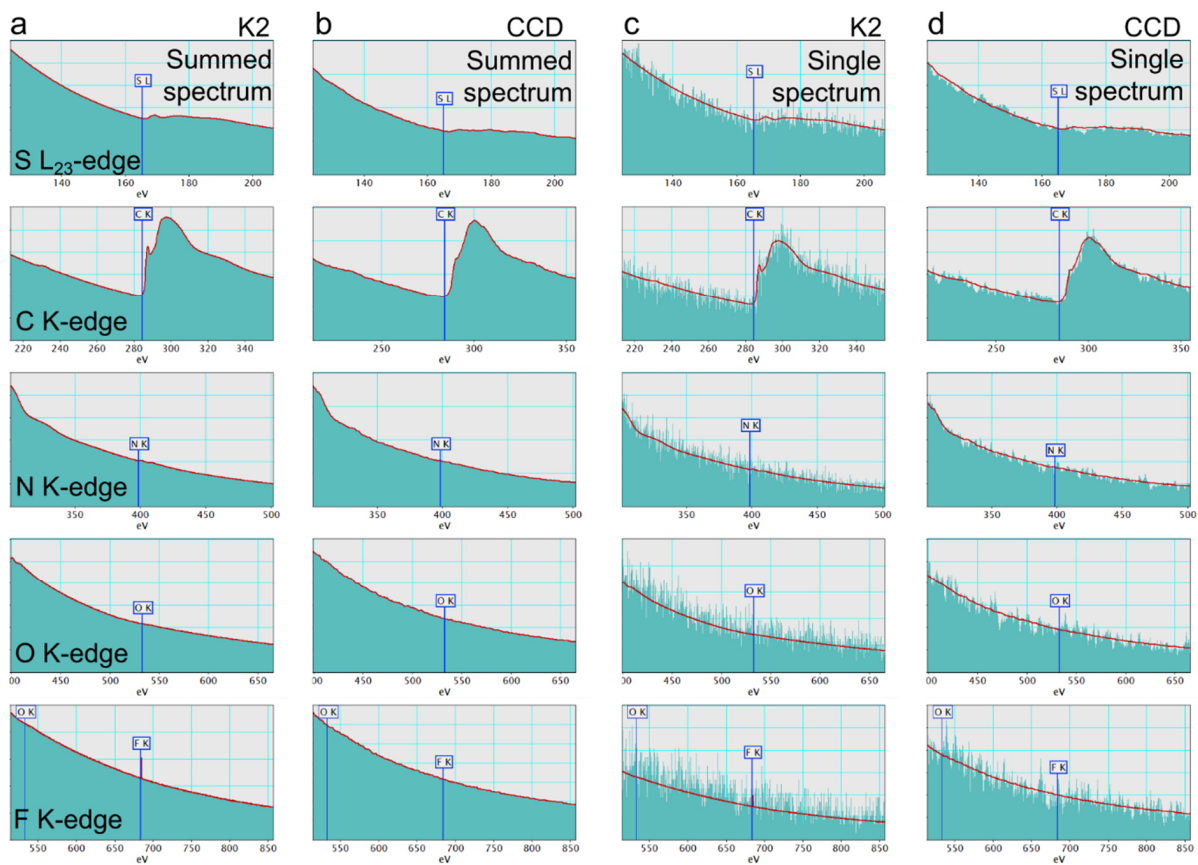

Fig. 2 Comparison of EELS data acquired with a direct electron detector (K2) and a conventional CCD camera (US1000). (a) Spectra summed over a full spectrum image for the K2 camera and (b) for the CCD camera. (c) Single pixel spectra recorded with the K2 camera and (d) the CCD camera. For all plots the original spectra are shown in green, while PCA treated spectra are shown in red.

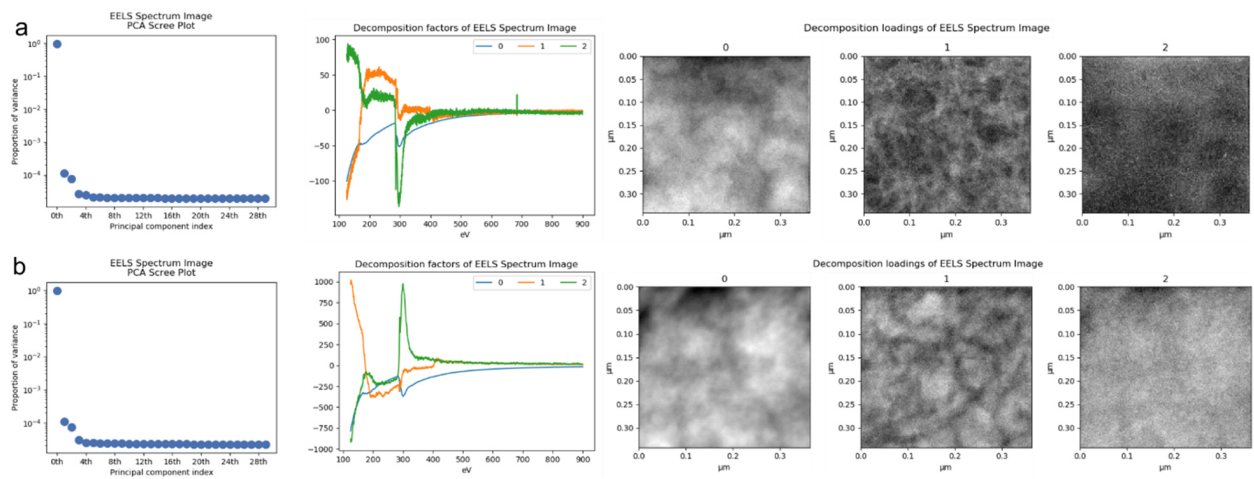

Fig. 3 PCA of a spectrum image acquired (a) on the K2 camera and (b) on the CCD camera, showing the scree plot, along with the decomposition loadings and factors of the first three principal components.

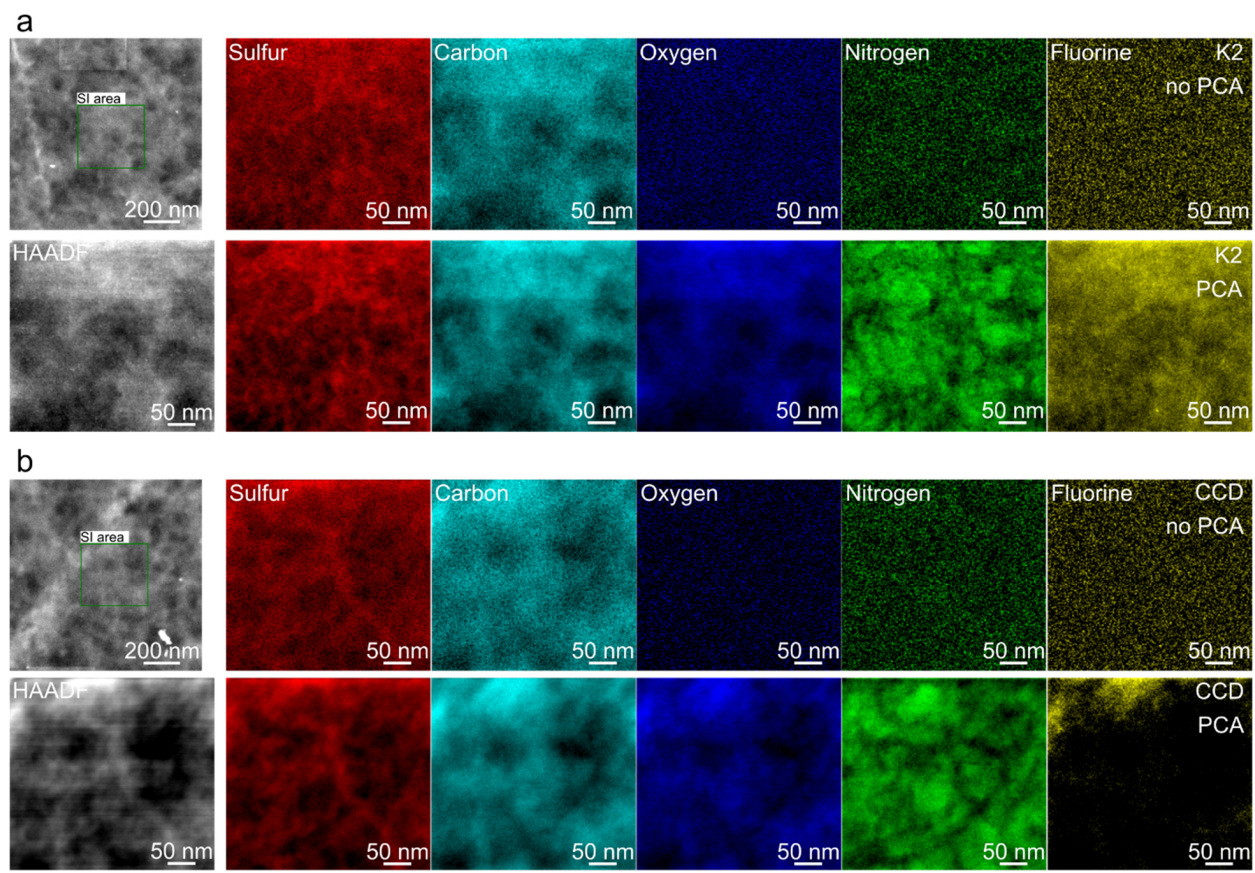

Fig. 4. Acquired HAADF STEM images and extracted elemental maps without and with PCA treatment on data (a) from the K2 camera and (b) from the CCD camera.

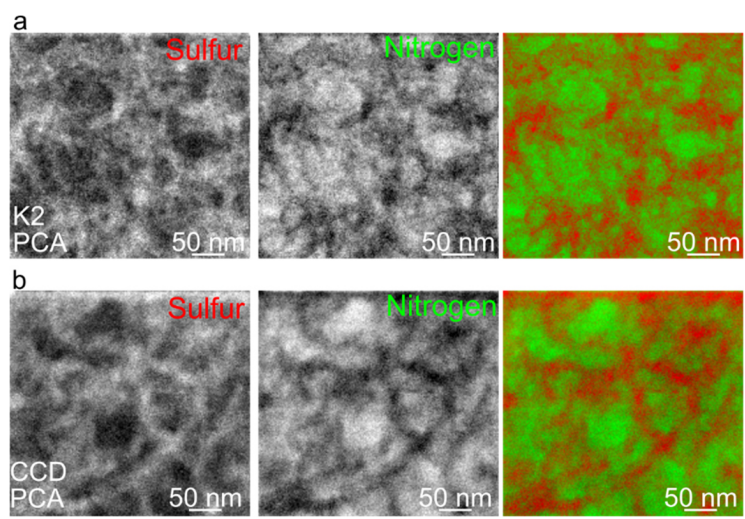

Fig. 5. Relative composition maps of sulfur and nitrogen and overlay (sulfur: red, nitrogen: green) for PCA treated data (a) from the K2 camera and (b) from the CCD camera.